# Half-metallic ferromagnetism in transition-metal doped germanium nitride: A first-principles study

Sheng-Li Zhang,[a,b] Wei Wang,[a,*] and Er-Hu Zhang[a]

[a]*Department of Applied Physics, Xi'an Jiaotong University, Xi'an 710049, China*

[b]*MOE Key Laboratory for Nonequilibrium Synthesis and Modulation of Condensed Matter, Xi'an Jiaotong University, Xi'an 710049, China*

The electronic and magnetic properties of transition-metal doped $\beta$-$Ge_3N_4$ have been studied using first-principles calculations. The results show that the substitutional transition-metal impurities tend to cluster. The V and Cr doped $Ge_3N_4$ compounds are ferromagnetic semiconductors, while the compounds with Mn and Fe doping show a half-metallic ferromagnetic character.

* Corresponding author at: Department of Applied Physics, School of Science, Xi'an Jiaotong University, Xi'an 710049, China. Tel./fax: +86 29 82664736.

E-mail address: wonvein@ymail.com

## I. INTRODUCTION

In the field of spintronics, many new materials are under investigation with the goal of achieving integrated electronic, magnetic and optic functionalities in a single material, leading to low cost, high-speed, small size and low power devices. These include diluted magnetic semiconductors (DMSs) which are synthesized by adding small amounts of transition-metal (TM) atoms into nonmagnetic semiconductors to obtain the composite materials with both ferromagnetic (FM) and semiconducting or half-metallic properties. The half-metal is the extreme case, in which one spin channel is metallic while the other spin one is strictly insulating. It is a sort of ideal material to design spin injection devices due to the 100% spin polarization. Early studies of DMSs are focused on II-VI semiconductors, such as CdSe and ZnTe [1]. Since many TM adopt divalent ionic states, the TM impurities substitute for divalent cations easily. However, most II-VI DMSs are antiferromagnetic (AFM), or have low Curie temperatures $T_C$. This renders them unattractive for application. The DMSs based on Mn doped III-V [2] and group IV [3] are also with lower $T_C$ (~150 K) and still difficult for application. Many attempts have been made to search DMSs with the $T_C$ higher than room temperature. Recently, some experimental reports indicated that the DMSs with Curie temperature higher than 300 K mainly concentrate in various TM doped oxides [4,5,6,7,8,9,10,11] and nitrides [12,13,14,15]. Unfortunately, the *p*-type doping in these compounds is very difficult [16]. It is necessary to keep up synthesizing a new class of room-temperature FM semiconductor or half-metal with *p*-type conductivity. Dietl *et al.* [17] calculated the magnetic properties of Mn doped various semiconductors, and concluded that wide band-gap semiconductors might offer better possibility for $T_C$ above room temperature based on the *p-d* exchange Zener model [18]. Therefore, the current interesting in the achievement of FM above room temperature has lead materials scientists mainly to focus on the wide band-gap semiconductors.

Germanium nitride ($Ge_3N_4$) is a potential wide band-gap semiconductor with excellent electro-optic properties. It at least has five possible structures, i.e., α, β, γ, pseudocubic, and graphitic phase. The β phase is the most stable one among these five possible structures [19,20,21]. It is a direct band-gap semiconductor with theoretical band-gap of 2.45 eV by density functional theory calculations with generalized gradient approximation (GGA) [20]. I. Chambouleyron *et al.* reported the experimental gap of *β* phase to be near 4.50 eV [22]. The first-principles calculation, which is one state of the art method to obtain an understanding of the electronic structure of materials, can guide us to discover new excellent materials. In this letter we performed systematic computational study on electronic and magnetic properties of β-$Ge_3N_4$ doped by TM (V, Cr, Mn and Fe) impurities, with the aim of expecting to discover promising compounds with the $T_C$ above room temperature. Furthermore, β-$Ge_3N_4$ has small lattice mismatch with Ge, thus the DMS devices based on β-$Ge_3N_4$ might be integrated with Ge-based devices. We focus on the single impurities doped β-$Ge_3N_4$, and study in detailed the structural, electronic, and magnetic properties of these systems as a function of TM impurities. Our results show that the TM impurities have tendencies to cluster together creating impurity rich regions. The β-$Ge_3N_4$ compounds with V and Cr doping show FM semiconducting behaviors, while the Mn and Fe doped compounds are found to display half-metallic FM characters. The remainder of this paper is organized as follows. In Sec. II, the details of the computational method are described. In Sec. III, we present our calculated results on structural, electronic and magnetic properties of $Ge_3N_4$ with TM doping. Finally, we give a short summary in Sec. IV.

## II. COMPUTATIONAL METHODLOGY

Our density functional theory calculations were performed using Vienna Ab initio Simulation Package (VASP) [23,24]. The projector augmented wave (PAW) method was used to describe the electron-ion interaction [25]. The exchange correlation between electrons was treated with GGA in the Perdew-Burke-Ernzerhof (PBE) form [26]. We used a cutoff energy of 450 eV for the plane-wave basis. The TM 3*d* and 4*s*, the Nitrogen 2*s* and 2*p*, and Germanium 4*s* and 4*d* electrons were treated as valence electrons.

The optimized lattice constants *a* and *c* of $Ge_3N_4$ primitive unit cell are 8.18 Å and 3.13 Å respectively, in good agreement with experiment [27]. For TM-doped calculations, we employed a supercell containing 3 primitive unit cells ($1a \times 1a \times 3c$). Two Ge atoms were substituted by two TM atoms in such a 42-atom supercell, which corresponds to a TM concentration of 11.11% roughly for the entire crystal. We explored two spatial arrangements, i.e., the near configuration (two TM atoms separated by one nitrogen atom) and the far one (two TM atoms separated by two -N-Ge-N-). In doing so, the first Mn atom was put at the site labeled by 1, and the second Mn atom was placed at site 2 or 3 (see Fig. 1) with the Mn-Mn distance being 3.13 Å for near spatial configuration and 6.03 Å for far spatial configuration. For all geometry optimizations, we kept *a* and *c* to the optimized values and relaxed only the internal freedom until the Hellmann-Feynman forces on all the atoms were converged less than 0.02 eV·Å$^{-1}$. A (4×4×3) mesh within Monkhorst-Pack scheme [28] was employed to sample the Brillouin-zone of the supercell and Gaussian smearing of 0.05 eV was applied to the Brillouin-zone integrations in total-energy calculations. For the density of states calculation, a (6×6×5) mesh within tetrahedron method [29] was adopted.

**Fig. 1.**

**III. RESULTS AND DISCUSSION**

Whether TM (from V to Fe) impurity is doped easily or not depends on its formation energy in Ge$_3$N$_4$. Thus, we start with examining the formation energy of a single TM ion occupying a substitutional site or an interstitial one in a 42-atom supercell (Mn content: 5.56%). The substitutional site and the optimized interstitial one (0.14$a$ 0.56$a$ 0.25$c$) are displayed in Fig. 1. The formation energy can be calculated by the formula $\Delta_F$= ($E_{TM}$+ $n\mu_{Ge}$)-($E_{pure}$+ $\mu_{TM}$), where $\mu_{Ge}$ and $\mu_{TM}$ are the chemical potential of bulk Ge and TM, and $n$=1, 0 for the substitutional and interstitial sites, respectively. In this process, we take the TM dopants as neutral defects, and the chemical potential of TM is obtained from the most stable structure of each TM, i.e., nonmagnetic bcc V, antiferromagnetic bcc Cr, antiferromagnetic fcc Mn and ferromagnetic bcc Fe.

**Table 1**

From the calculated results presented in Table 1, it is found that the substitutional site is more energetically favorable than the interstitial one by 5.97 eV, 5.39 eV, 3.88 eV and 2.86 eV for V, Cr, Mn and Fe occupying respectively. This means that the all TM impurities are expected to more likely occupy substitutional site rather than interstitial one and the probability of TM ions residing on the interstitial sites may be very small. Moreover, it is noted that the heat formation of substitutional TM increases from -1.15 eV for V to +1.89 eV for Fe on moving right through the 3$d$ series. This implies that the solid solubility of TM is expected to lower gradually as the doptant is replaced from V to Fe. In other words, the V (Fe) impurity can be most easily (difficultly) doped in Ge$_3$N$_4$. It should be pointed out that the formation energy also varies with the TM content, but we believe a study of formation energy of substitutional and interstitial TM as a function of TM series at a TM content of 5.56% could shed some light on understanding the thermodynamics tendency of TM doped $Ge_3N_4$ compounds.

In order to investigate the thermodynamic tendency of distribution and magnetic coupling of TM impurities, we relaxed all internal coordinates for both AFM and FM orderings in each spatial configuration. In Fig. 2 we show that the calculated total energies of all configurations in reference to the corresponding FM ordering of near configuration (set to zero and not shown). If we define the total energy difference between near and far configurations as segregation energy (a negative value implies that clustering is favorable). We find that the calculated segregation energy is -0.04 eV, -0.11 eV, -0.78 and -0.46 eV for the compounds with V, Cr, Mn and Fe doping respectively. This suggests that the TM (from V to Fe) impurities cluster together during growth, rather than distribute uniformly throughout the entire crystal.

As far as the magnetic coupling between TM ions is concerned, we can see that the FM state is more strongly favorable for the near configuration and the AFM state is slightly more favorable for the far configuration for all systems, except for the compound with Fe doping which always shows FM state for both spatial configurations (see Fig. 2). In consideration of fact that all compounds with TM doping prefer near configuration, we define the FM stabilization energy as the total energy difference between FM and AFM orderings of near configuration, $\Delta E=E_{FM}-E_{AFM}$ (a negative $\Delta E$ implies that FM is favorable). The calculated FM stability energy $\Delta E$ is -0.04 eV, -0.09 eV, -0.17 eV and -0.09 eV for the compounds with V, Cr, Mn and Fe doping respectively. It is clear that all these compounds show FM ground states and have strong FM coupling (especially the Mn doped system). A stronger FM coupling implies that a higher Curie temperature $T_C$ might be reached. The Curie temperature $T_C$ can be estimated roughly based on the mean-field approximation, and calculated according to the following expression [30,31,32],

$$T_C = \frac{2}{3k_B} \cdot \Delta \quad (1)$$

Where $\triangle$ is the total energy difference between the spin-glass and FM state per unit cell, and $k_B$ denotes Boltzmann constant. Here the $\triangle$ can also be well approximately chose to be the total energy difference between AFM and FM phases. Using Eq. (1), we estimated that the $T_C$ is 309 K, 696 K, 1314 K and 696 K at a TM content of 11.11% with a clustering distribution for the V, Cr, Mn and Fe doped $Ge_3N_4$-based systems respectively. It is evident that the TM doped $Ge_3N_4$ compounds have the calculated Curie temperatures reaching to room temperature; especially the Curie temperature of Mn doped compound exceeds 1000 K. We believe that many TM clusters present in real TM doped $Ge_3N_4$ samples, and the $\triangle$ may be expected decrease exponentially with respect to the growing cluster-cluster distance. Therefore the mean-field approximation used here may be expected to give an upper limit in estimating $T_C$.

**Fig. 2.**

To investigate the role of TM in magnetic coupling interactions, we summarize the total and local magnetic moments (obtained by integrating up to a Wigner-Seitz radius of 1.1 Å) for TM doped $Ge_3N_4$ systems. As displayed in Table 2, the stable magnetic moment is $1\mu_B$, $2\mu_B$, $3\mu_B$ and $4\mu_B$ per V, Cr, Mn and Fe ion respectively. The integer total magnetic moments imply that these compounds would have semiconducting or half-metallic characters, which will be verified by the calculated density of states in the following discussion. Furthermore, it is also found that the total magnetic moments are mainly derived from TM $3d$ states (see Table 2), giving further evidence of the strong localized character of spin density distribution in DMSs. The nearby Ge atoms contribute negligible positive spin moments, and the N atoms also give a very small negative contribution (less than $0.1\mu_B$ per N) which increases as the number of nearest-neighboring TM increases.

**Table 2**

In Fig. 3, we plot the calculated total density of states (TDOS) for these compounds. In all cases the TDOS for FM orderings are shown. The majority spin states are plotted along the positive y direction, and the minority states are plotted along the negative y direction. The Fermi energy is set to zero in energy scale. From the TDOS results, we find that the minority states always keep semiconducting characters for all TM (from V to Fe) impurities doped $Ge_3N_4$ systems. Compared with minority states, the majority states dramatically vary with TM dopants. For the compound with V doping, the band gap of majority states seems to be zero. This is likely an artifact to the underestimation of the band gap by the GGA. Thus, the V doped $Ge_3N_4$ should be classified as a semiconductor. The majority states of system with Cr doping shows clearly semiconducting behavior, while the majority states of Mn and Fe doped β-$Ge_3N_4$ compounds have a half-metallic character.

**Fig. 3.**

In order to get an insight into the effects of TM on electronic structures of TM doped $Ge_3N_4$ compounds, we display the TM $3d$ projected density of states (PDOS) in Fig. 4. It is noted that the region between -6 eV and -2 eV in PDOS is derived from the N $2p$ states (not shown). The TM $3d$ states are within the same energy window of DOS and can hybridize strongly with N $2p$ states, which results in a broad band with a width comparable to that of the N $p$ band for all compounds. Moreover, the region between -2 eV and 0 eV near Fermi lever is mainly derived from the TM $3d$ states. Therefore the electronic properties of these compounds are dominantly determined by the TM $3d$ states. For V and Cr, the highest occupied states of the majority component are just below the Fermi level and fully occupied. In the Mn or Fe doped cases, a finite "acceptor" state is observed to cross the Fermi level, making the majority state shows $p$ type character. In contrast with the majority states, the minority $3d$ states of all TM impurities are always deep in valence band and keep semiconducting characters. Because of these different features of TM 3*d* states, the electronic properties of the compounds with various TM dopants are different from each other. The occupation of majority (minority) component is 1.98 (0.98), 2.69 (1.07), 3.82 (1.07) and 4.40 (1.34) in the atomic sphere with a radius of 1.1 Å for V, Cr, Mn and Fe respectively. Such a variety of electron occupation should be also reflected in the PDOS. As expected, it is found that the energy of the majority states clearly shift down relative to the Fermi level in energy, while the minority states have little changes as the TM series move from V to Fe.

**Fig. 4.**

**V. SUMMARY**

In summary, we have performed a systematic study of the structural, electronic and magnetic properties of TM (V, Cr, Mn and Fe) doped β-$Ge_3N_4$ using first principles calculations. We studied the formation energies of substituional TM and interstitial one at a TM content of 5.56%. The results show that the substituional site is more energetically favorable for TM occupying. Furthermore, the solid solubility of TM is expected to lower gradually as the doptant is replaced from V to Fe. We also studied the thermodynamic tendency of TM impurities by investigating two spatial doped configurations: near configuration and far configuration. It is found that all impurities have a tendency to occupy nearest-neighbor substituional sites. All compounds with TM doping show a FM ground state and strong FM coupling with a clustering distribution. We estimated the Curie temperature $T_C$ based on mean-field approximation and found that the calculated $T_C$ of these compounds with TM doping reach or above room temperature. Finally, we investigated the electronic structure of TM doped $Ge_3N_4$ with a clustering distribution. The Mn and Fe doped $Ge_3N_4$ compounds show a half-metallic property, while the $Ge_3N_4$ systems with V and Cr doping display a semiconducting behavior. In view of half-metallic FM character with the $T_C$ higher than 1000 K and $p$-type conductivity, the best candidate to realize a magnetic semiconductor based on $Ge_3N_4$ is Mn dopant. Therefore, the $Ge_3N_4$-based compounds with Mn doping could be potentially significant materials for spin-injection applications for spintronics devices.

**ACKOWLEDGMENTS**

We thank Prof. Wen-Tong Geng for help in design, discussion, and manuscript preparation for this work. We also thank Xue-Guang Xu for valuable suggestion. This work was supported by the Cultivation Fund of the Key Scientific and Technical Innovation Project，Ministry of Education of China (NO 708082). The Calculations were performed on console1 of XJTU.

Table 1. The calculated formation energy $\Delta_F$ (eV/TM) for a single TM defect occupying a substitutional site (TM$^{Sub}$) or an interstitial one (TM$^{Int}$) as a function of the TM series in $\beta$-Ge$_3$N$_4$.

| Dopant | V | Cr | Mn | Fe |
|---|---|---|---|---|
| $\Delta_F$ of TM$^{Sut}$ (eV/TM) | -1.15 | 0.08 | 0.12 | 1.89 |

| $\Delta_F$ of TM$^{Int}$ (eV/TM) | 4.82 | 5.47 | 4.00 | 4.75 |

Table 2. The calculated total magnetic moments $\mu_{tot}$ ($\mu_B$/cell) for a whole unit cell and the local average magnetic moments $\mu_{TM}$ ($\mu_B$/TM) in a atomic sphere of 1.10 Å.

| | TM | V | Cr | Mn | Fe |
|---|---|---|---|---|---|
| GGA | $\mu_{tot}$ ($\mu_B$/cell) | 2.00 | 4.00 | 6.00 | 8.00 |
| | $\mu_{TM}$ ($\mu_B$/TM) | 0.94 | 1.83 | 2.59 | 3.05 |

**Figure Captions**

**Fig. 1. (Color online) The supercell used to model TM-doped β-Ge$_3$N$_4$. Orange and silver balls represent Ge and N respectively.**

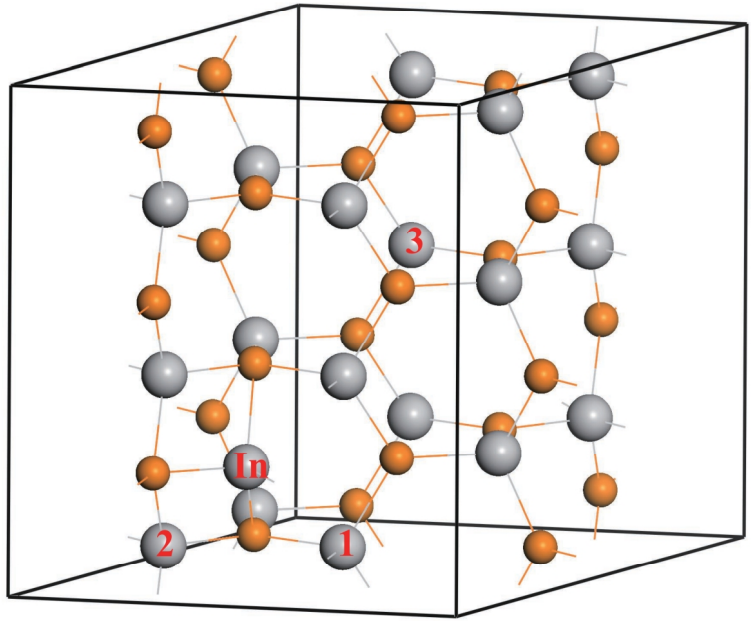

**Fig. 2.** (Color online) The calculated total energies of FM and AFM spin orderings in reference to the corresponding FM alignment of near spatial configuration (set to zero and not shown) as a function of TM-TM distance.

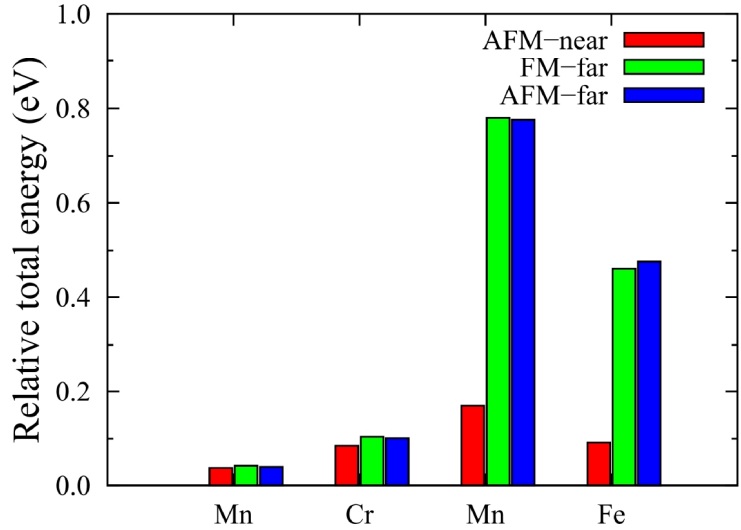

**Fig. 3.** The TDOS of FM orderings for TM doped β-Ge$_3$N$_4$ systems. Positive (negative) values refer to majority (minority) spin component. The zero of the energy scale is set at the Fermi energy.

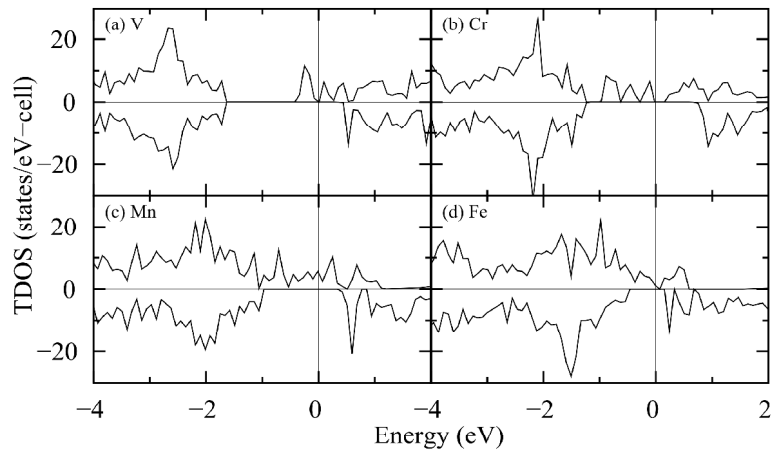

**Fig. 4. The calculated 3*d* PDOS of TM impurities. Positive (negative) values refer to majority (minority) spin component. The zero of the energy scale is set at the Fermi energy.**

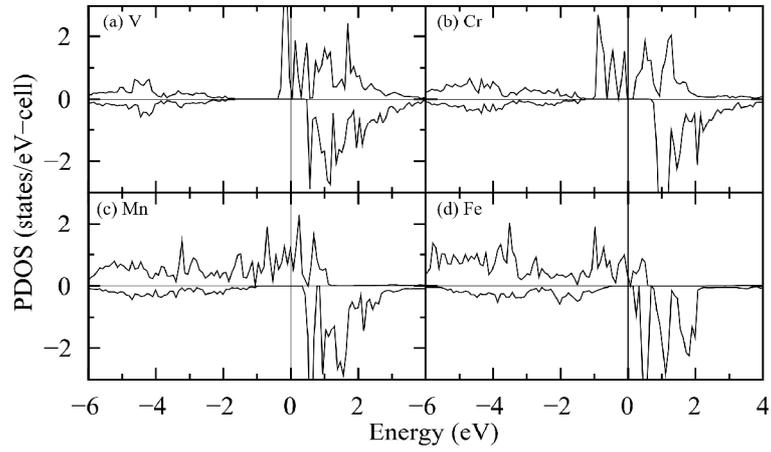